\documentclass[12pt]{iopart}

\usepackage{amsfonts}
\usepackage{multirow}
\usepackage{graphicx}
\eqnobysec

\usepackage{soul}

\usepackage{color}
\definecolor{darkgreen}{rgb}{0,0.6,0}

\newcommand{\misc}[1]{}
\begin{document}

\title[Bipartite Entanglement, Partial Transposition and the Uncertainty Principle]{Bipartite Entanglement, Partial Transposition and the Uncertainty Principle for Finite-Dimensional Hilbert Spaces}

%%%%%%%%%%%%%%%%%%%%%%%%%%%%%%%%%%%%%%%%%%%%%%%%
\author{Yehuda B. Band$^1$ and Pier A. Mello$^2$}

\address{$^1$Department of Chemistry, Department of Electro-optics,
Department of Physics, 
and The Ilse Katz Institute for Nanoscale Science and Technology,
Ben-Gurion University, Beer Sheva, Israel, 84105, and\\
New York University and the NYU-ECNU Institute of Physics at NYU Shanghai, 3663 Zhongshan Road North, Shanghai, 200062, China}

\address{$^2$Instituto de F\'{\i}sica, Universidad Nacional Aut\'{o}noma de M\'{e}xico, Apartado Postal 20-364, 01000 M\'{e}xico, D.F., Mexico}

\ead{mello@fisica.unam.mx,band@bgu.ac.il}

%%%%%%%%%%%%%%%%%%%%%%%%%%%%%%%%%
\begin{abstract}
We first show that partial transposition for pure and mixed two-particle states in a discrete $N$-dimensional Hilbert space is equivalent to a change in sign of the momentum of one of the particles in the Wigner function for the state.  We then show that it is possible to formulate an uncertainty relation for two-particle Hermitian operators constructed in terms of Schwinger operators, and study its role in detecting entanglement in a two-particle state: the violation of the uncertainty relation for a partially transposed state implies that the original state is entangled.  This generalizes a result obtained for continuous-variable systems to the discrete-variable-system case.  This is significant because testing entanglement in terms of an uncertainty relation has a physically appealing interpretation. We study the case of a Werner state, which is a mixed state constructed as a convex combination with a parameter $r$ of a Bell state $|\Phi^{+} \rangle$ and the completely incoherent state, $\hat{\rho}_r = r |\Phi^{+} \rangle \langle \Phi^{+}| + (1-r)\frac{\hat{\mathbb{I}}}{N^2}$: we find that for $r_0 < r < 1$, where $r_0$ is obtained as a function of the dimensionality $N$, the uncertainty relation for the partially transposed Werner state is violated and the original Werner state is entangled.  
\end{abstract}
%%%%%%%%%%%%%%%%%%%%%%%%%%%%%%%%%

% 03.67.-a Quantum information
% 03.67.Mn Entanglement measures, witnesses, and other characterizations
% 03.65.Ud Entanglement and quantum nonlocality

% \vspace{0.5cm}
% 
% \hspace{4cm}Version *band-mello-final*
% \hspace{1cm}Tue, Jan.~12, 1~pm, 2016

\maketitle

%%%%%%%%%%%%%%%%%%%%%%%%%%%%%
\section{Introduction}  \label{intro}

Quantum entanglement occurring in multipartite qubit (and qunits, i.e., $N$-dimensional quantum bits) states is a powerful computation and information resource \cite{Steane_98, band-avishai}.  Entanglement of pure quantum states is well understood, but entanglement of mixed quantum states, states that cannot be represented using a wave function but must be described using a density matrix, is not yet fully understood.  For pure bipartite states, Schmidt coefficients relate the degree of entanglement to the von Neumann entropy of the reduced density matrix associated with either of the two subsystems; a pure state with vanishing von Neumann entropy corresponds to a separable state, whereas one with finite von Neumann entropy is entangled, and one with maximum von Neumann entropy is maximally entangled.  But no general measure of entanglement of mixed states exists.  Even deciding whether a mixed state is entangled or not is not always an easy task for mixed states.  A large variety of measures have been studied in the literature to quantify entanglement for a given state, as discussed in Ref.~\cite{plenio_virmani_2006}.  Entanglement witnesses, tools used to determine whether a state is separable or not, have been proposed \cite{plenio_virmani_2006}.  A useful concept in this context is {\em partial transposition} (PT) with respect to one of the particles \cite{peres_96,horodecki_97,horodecki_98}: when the partially transposed state is not a legitimate quantum mechanical (QM) state, the original state is entangled.

It has been noted in Ref.~\cite{simon_2000} (see also Refs.~\cite{werner-wolf2001, braunstein_van_loock_2005}) that, for continuous variables, partial transposition of one particle of a bipartite state amounts to a change in sign of the momentum of that particle in the Wigner function (WF) of the state.  Various uncertainty relations which must be fulfilled for a legitimate QM state were studied for continuous variables \cite{simon_2000,duan_et_al_2000}, and conditions were found for their validity in a PT state, their violation implying entanglement of the original state.

These results for continuous variables can be illustrated in a particularly clear fashion in the special case of the ``normalized version" of the EPR pure state,
%%%%%%%%%%%%%%%%%%%
\begin{equation}  \label{EPR normalizable}
\Psi(x,X)= \frac{e^{-\frac{(x-d)^2}{4\sigma_x^2}}}{(2\pi \sigma_x^2)^{-1/4}} 
\cdot 
\frac{e^{-\frac{X^2}{4\sigma_X^2}}}{(2\pi \sigma_X^2)^{-1/4}} \, ,
\end{equation}
%%%%%%%%%%%%%%%%%%%
where $x$ and $X$ are the relative and center-of-mass coordinates of the two particles.  
Partial transposition for particle 1  ($T_1$) in the coordinate basis is defined by the transformation \cite{peres_96,horodecki_97,horodecki_98}
%%%%%%%%%%%%%%%%%%%
\begin{equation}
\rho (q_1, q_2; q'_1, q'_2) 
\rightarrow \rho^{T_1} (q_1, q_2; q'_1, q'_2)  
\equiv \rho (q'_1, q_2; q_1, q'_2) \; .
\label{PT in q basis}
\end{equation}
%%%%%%%%%%%%%%%%%%%
Here, $q_i$ is the coordinate of particle $i$ and $\hat{\rho}$ is the density operator of the system.  If the resulting $\hat{\rho}^{T_1}$ is not a legitimate quantum mechanical state, the original state $\hat{\rho}$ is entangled.
The WF is affected by $T_1$ in the coordinate basis as 
%%%%%%%%%%%%%%%%%%%
\begin{equation}
W_{\hat{\rho}^{T_1}}(q_1,q_2;p_1,p_2)
=W_{\hat{\rho}}(q_1,q_2;-p_1,p_2) \; ,
\label{PT and W cont}
\end{equation}
%%%%%%%%%%%%%%%%%%%ta
thus changing the sign of momentum $p_1$.
The WF of state (\ref{EPR normalizable}) is
%%%%%%%%%%%%%%%%%%%
\begin{equation}
W_{\hat{\rho}}(q,Q;p,P)
= 2e^{-\frac{(q-d)^2}{2\sigma_x^2}}e^{-2\sigma_x^2 p^2}
\cdot 2e^{-\frac{Q^2}{2\sigma_X^2}}e^{-2\sigma_X^2 P^2} \; ,
\label{WF of rho 1}
\end{equation}
%%%%%%%%%%%%%%%%%%%
or, in terms of the individual-particle positions and momenta $q_1,q_2,p_1,p_2$,
%%%%%%%%%%%%%%%%%%%
\begin{equation}
W_{\hat{\rho}}(q_1,q_2;p_1,p_2)
= 4e^{-\frac{(q_1-q_2-d)^2}{2\sigma_x^2}}e^{-2\sigma_x^2 (\frac{p_1-p_2}{2})^2}
\cdot e^{-\frac{\left(\frac{q_1+q_2}{2}\right)^2}{2\sigma_X^2}}e^{-2\sigma_X^2 (p_1+p_2)^2}  .
\label{WF of rho 2}
\end{equation}
%%%%%%%%%%%%%%%%%%%
The WF resulting from the $T_1$ operation is, according to 
Eq. (\ref{PT and W cont}), given by
%%%%%%%%%%%%%%%%%%%
\begin{equation}
W_{\hat{\rho}^{T_1}}(q_1,q_2;p_1,p_2)
= 4e^{-\frac{(q_1-q_2-d)^2}{2\sigma_x^2}}
e^{-2\sigma_x^2 (\frac{p_1+p_2}{2})^2}
\cdot e^{-\frac{\left(\frac{q_1+q_2}{2}\right)^2}{2\sigma_X^2}}
e^{-2\sigma_X^2 (p_1-p_2)^2} \;.
\label{WF of sigma 1}
\end{equation}
%%%%%%%%%%%%%%%%%%%
The variances of $q,p,Q$, and $P$ for $\hat{\rho^{T_1}}$ are
%%%%%%%%%%%%%%%%%%%
\numparts
\begin{eqnarray}
[(\Delta q)^2]_{\hat{\rho}^{T_1}} &=& \sigma_x^2 \; , \hspace{13mm} 
[(\Delta p)^2]_{\hat{\rho}^{T_1}} = \frac{1}{16 \sigma_X^2} \; , \\
\left[ (\Delta Q)^2 \right]_{\hat{\rho}^{T_1}} &=& \sigma_X^2 \; , \hspace{1cm} 
\left[(\Delta P)^2 \right]_{\hat{\rho}^{T_1}} = \frac{1}{\sigma_x^2} \; .
\label{Delta q, etc for sigma}
\end{eqnarray}
\endnumparts
%%%%%%%%%%%%%%%%%
Hence, $[(\Delta q)^2]_{\hat{\rho}^{T_1}} [(\Delta p)^2]_{\hat{\rho}^{T_1}}$ is arbitrary and {\em may violate the uncertainty principle};  similarly for $[(\Delta Q)^2]_{\hat{\rho}^{T_1}} [(\Delta P)^2]_{\hat{\rho}^{T_1}}$: if the relative canonical variables violate the uncertainty principle, the center-of-mass ones do not, and {\it vice versa}.  
If $\sigma_x^2/\sigma_X^2 = 4$, 
the product of these uncertainties is $1/4$
%$[(\Delta q)^2]_{\hat{\rho^{T_1}}} [(\Delta p)^2]_{\hat{\rho^{T_1}}}= [(\Delta %Q)^2]_{\hat{\rho^{T_1}}} [(\Delta P)^2]_{\hat{\rho^{T_1}}}= 1/4$, 
and the uncertainty principle is not violated.
Thus, except for this exceptional case, PT on the entangled pure state (\ref{EPR normalizable}) does not lead to a physical state, as {\em it violates the uncertainty principle}: this verifies the fact that the original state (\ref{EPR normalizable}) is entangled.

In the above continuous variable example, the violation of the uncertainty principle is a clear consequence of partial transposition of the entangled state in question.  Extending the above example to discrete-variable systems defined in a finite-dimensional Hilbert space turns out to be of real value but is non-trivial: this extension is the purpose of the present paper.  The extension is facilitated by the generalization of the concept of Wigner function to the discrete variable case, which has been widely studied in the literature (see Ref.~\cite{mello_revzen_2014} and references cited therein).  Here, the formulation presented in Refs.~\cite{mello_revzen_2014, mann_mello_revzen_2015} is used to prove that, for the discrete case, PT can be interpreted in terms of a change in sign of the momentum of one of the particles in the Wigner function of the state.  Just as for for the continuous case, this statement is appealing, as it gives an intuitive interpretation of PT.
We find that the use of Schwinger operators \cite{schwinger} allows the generalization from the continuous-variable case to the finite-dimensional one.  Hermitian operators can be constructed using the Schwinger operators, and an uncertainty relation (UR) can be formulated to detect entanglement of two-particle states: such an UR must be fulfilled for a legitimate QM state.  Its violation for a PT `state' signals entanglement in the original state.  To the best of our knowledge, this analysis, originally carried out for continuous-variable systems, has not been pursued for the discrete-variable case: in our view, an UR provides a measure having a clear physical interpretation to test entanglement.

This paper is organized as follows.  In the next section we use Schwinger operators to formulate an UR for states defined in a discrete $N$-dimensional Hilbert space.  The procedure is first illustrated for one-particle states in Sec.~\ref{schwinger-1-part} and is then extended in Sec.~\ref{schwinger-2-parts.} to two-particle states.  In Sec.~\ref{schwinger-2-parts.} we study the effect of PT on the Wigner function of a two-particle state.  
In Sec.~\ref{applications} we study the application of the PT-UR criterion to a Werner state, which is a mixed state constructed as a convex combination with a parameter $r$ of a Bell state and the completely incoherent state, and find the values of $r$ for which the uncertainty relation for the PT state is violated and the original Werner state is entangled.  
Conclusions and a summary are presented in Sec.~\ref{conclu}.  Four appendices present a number of results mentioned in the text without interrupting the main flow of the presentation.

%%%%%%%%%%%%%%%%%%%%%%
\section{Schwinger operators and uncertainty relations}
\label{schwinger}

By way of introduction, we first consider just one particle that can be modeled in terms of a discrete $N$-dimensional Hilbert space.
We then extend the analysis to two-particle systems, which is the main topic of this paper.

%%%%%%%%%%%%%%%%%%%%%%
\subsection{One particle}
\label{schwinger-1-part}

Consider a one-particle system with with a discrete, finite set of states.  The eigenvalues of observable operators take on a discrete set of values and the quantum description is given in terms of a finite-dimensional Hilbert space.  As an example, consider a system with angular momentum $j$, described in a Hilbert space of dimensionality $2j+1$.  Another example is the position and momentum observables taken on a discrete lattice of finite dimensionality $N$ (see, e.g., Ref.~\cite{de_la_torre-goyeneche}).  The latter case is the one we shall be concerned with in this paper.

The Hilbert space to be considered is thus spanned by $N$ distinct states $|q\rangle$, with $q=0, 1, \cdots, (N-1)$.  As discussed in \ref{schwinger-one-particle}, the periodicity condition $|q+N\rangle = |q\rangle$ is imposed.  The Schwinger operators \cite{schwinger} $\hat{X}$ and $\hat{Z}$ are also defined in \ref{schwinger-one-particle}, as are the operators $\hat{q}$ and $\hat{p}$.  
As $\hat{X}$ {\em performs translations in the variable $q$ and $\hat{Z}$ 
in the variable $p$, we designate $\hat{q}$ and $\hat{p}$ as the position and momentum operators}.  
Notice, however, that their commutation relation for finite $N$ is quite complicated [e.g., see Ref.~\cite{de_la_torre-goyeneche}, Eq.~(20)], and that in the continuous limit their commutator reduces to the standard form 
\cite{de_la_torre-goyeneche, durt_et_al}, $[\hat{q}, \hat{p}] = i$.

\ref{property of P(p1,p2)} shows that under partial transposition of the density matrix in the coordinate representation and for $N>2$ (for $N=2$, $| p\rangle = |-p\rangle$), the probability distribution of momentum is affected as follows:
%%%%%%%%%%%%%%%%%
\begin{equation}   \label{P(p) 1part}
P_{\rho^{T}}(p) = P_{\rho}(-p).
\end{equation}
%%%%%%%%%%%%%%%%%
Thus, transposition in the {\em coordinate representation} has the intuitive physical meaning  of changing the sign of momentum $p$ in the momentum probability distribution, an effect which corresponds to {\em time-reversal} (if no spin is present).  Moreover, the Wigner function defined as in Refs.~\cite{mello_revzen_2014, mann_mello_revzen_2015} has the property,
%%%%%%%%%%%%%%%%%%%
\begin{equation}
W_{\hat{\rho}^{T}}(q,p)
=W_{\hat{\rho}}(q,-p) \; ,
\label{PT and W discrete 1part}
\end{equation}
%%%%%%%%%%%%%%%%%%%
as demonstrated in \ref{proof of PT and W discrete}, thus again exhibiting a {\em change in sign of} $p$.  The definition of the Wigner function of Refs.~\cite{mello_revzen_2014, mann_mello_revzen_2015} requires $N$ to be a prime number larger than 2.  It turns out that this is the simplest extension of the continuous case to the discrete one that one can study, which can then be extended to the case where $N$ is not prime (see, e.g., Ref.~\cite{wootters87,wooters-fields89}).  When $N$ is a prime number, the integers $0, \cdots, N-1$ form a mathematical field playing a role parallel to that of the field of real numbers in the continuous case.  Also, in this case, a set of $N+1$ mutually unbiased basis states is known \cite{ivanovic36}.  In what follows, when the Wigner function is not involved, the prime dimensionality requirement is not needed.

The Hermitian operators
%%%%%%%%%%%%%%%%%%%%
\numparts
\begin{eqnarray}
\hat{A} &=& \frac{1}{2i}(Z-Z^{\dagger}) 
= \sin \frac{2 \pi \hat{q}}{N}
=\hat{A}^{\dagger} \; ,   
\label{A 1p} \\
\hat{B} &=& -\frac{1}{2i}(X-X^{\dagger})
= \sin \frac{2 \pi \hat{p}}{N}
=\hat{B}^{\dagger} \; ,
\label{B 1p}
\end{eqnarray}
\endnumparts
%%%%%%%%%%%%%%%%%%%%
which are physical observables, are formed using the Schwinger operators $\hat{X}$ and $\hat{Z}$ introduced in \ref{schwinger-one-particle}. 
For these Hermitian operators one finds the uncertainty relation (see, e.g., Ref.~\cite{ballentine}, Eq.~(8.31), p.~224)
%%%%%%%%%%%%%%%%%%%%%%
\begin{equation}
(\Delta \hat{A})_{\hat{\rho}}^2 \; (\Delta \hat{B})_{\hat{\rho}}^2 
\geq \frac14 \left|\left\langle [\hat{A}, \hat{B}] \right\rangle _{\hat{\rho}}\right|^2 ,
\label{unc-rel 1p 1}
\end{equation}
%%%%%%%%%%%%%%%%%%%%%%
where $\hat{\rho}$ is the density operator for a state of interest; the expectation values in (\ref{unc-rel 1p 1}) are to be evaluated with it.  To obtain this relation one requires the following properties for the density operator and the observables:
%%%%%%%%%%%%%%%%%%%%%%
\numparts
\begin{eqnarray}
\hat{\rho} &=& \hat{\rho}^{\dagger}, 
\label{rho Herm} \\
\hat{\rho} &\ge& 0, \;\; {\rm i.e.,}  \; \hat{\rho}  \; {\rm is \; nonnegative \; definite} ,
\label{rho >0} \\
\hat{A} &=& \hat{A}^{\dagger} ,
\label{A Herm} \\
\hat{B} &=& \hat{B}^{\dagger} .
\label{B Herm}
\end{eqnarray}
\label{requirments for unc. rel.}
\endnumparts
%%%%%%%%%%%%%%%%%%%%%%
Note that, quite generally, given a Hermitian operator $\hat{O}=\hat{O}^{\dagger}$, one may introduce the ``expectation value''
%%%%%%%%%%%%%%%%%%%%%%
\begin{equation}
\langle \hat{O} \rangle_{\hat{\pi}}
\equiv {\mathrm{Tr}}(\hat{\pi} \hat{O}) \; ,
\end{equation}
%%%%%%%%%%%%%%%%%%%%%%
where $\hat{\pi}$ may not necessarily qualify as a density operator; i.e., it may not fulfill
\numparts 
\begin{eqnarray}
\hat{\pi}&=&\hat{\pi}^{\dagger}, \;\; (\rm Hermiticity)
\label{pi a} \\
\hat{\pi} &\ge& 0, \;\; ({\rm nonnegative \; definiteness}) .
\label{pi b}
\end{eqnarray}
\endnumparts

The general idea in what follows is, for Hermitian operators $\hat{A}$ and $\hat{B}$, to put the relation
%%%%%%%%%%%%%%%%%%%%%%
\begin{equation}
(\Delta \hat{A})_{\hat{\pi}}^2 \; (\Delta \hat{B})_{\hat{\pi}}^2 
\ge \frac14 \left|\left\langle [\hat{A}, \hat{B}] \right\rangle _{\hat{\pi}}\right|^2 \;
\label{}
\end{equation}
%%%%%%%%%%%%%%%%%%%%%%
to the test.  If we find that this relation is violated, either (\ref{pi a}) or (\ref{pi b}), or both, are violated.

We return to $\hat{A}$ and $\hat{B}$ defined in Eqs.~(\ref{A 1p}) and (\ref{B 1p}). Explicitly computing the commutator $[\hat{A}, \hat{B}]$, we find
%%%%%%%%%%%%%%%%%%%%%%
\begin{eqnarray}  \label{[A,B]}
[\hat{A}, \hat{B}]
&=& \frac14 
\left\{
[\hat{Z}, \hat{X}] - [\hat{Z}, \hat{X}^{\dagger}] - [\hat{Z}^{\dagger}, \hat{X}] + [\hat{Z}^{\dagger}, \hat{X}^{\dagger}]
\right\} \nonumber  \\
&=& \frac14
\left[(\omega -1)\left(\hat{X}\hat{Z} + \hat{Z}^{\dagger}\hat{X}\right)
-(\omega^* -1)\left(\hat{Z}^{\dagger}\hat{X}^{\dagger} + \hat{X}^{\dagger}\hat{Z}\right)
\right] \; ,
\end{eqnarray}
%%%%%%%%%%%%%%%%%%%%%%
where $\omega$ is defined in Eq.~(\ref{Z}) of \ref{schwinger-one-particle}, i.e., $\omega=e^{2 \pi i/N}$.  For a general density-like operator $\hat{\pi}$, we check the relation 
%%%%%%%%%%%%%%%%%%%%%%
\begin{eqnarray}
(\Delta \hat{A})_{\hat{\pi}}^2 \; (\Delta \hat{B})_{\hat{\pi}}^2 
&\geq& \frac{1}{64}
\left| 
(\omega -1)\left(\langle \hat{X}\hat{Z}\rangle_{\hat{\pi}} + \langle \hat{Z}^{\dagger}\hat{X}\rangle_{\hat{\pi}} \right)
\right.
\nonumber \\
&& \;\;\;\;  \left. -(\omega^{*} -1)
\left(\langle \hat{X}^{\dagger}\hat{Z}\rangle_{\hat{\pi}} + \langle \hat{Z}^{\dagger}\hat{X}^{\dagger}\rangle_{\hat{\pi}} \right)
\right|^2 \; .
\label{unc-rel 1p 2}
\end{eqnarray}
%%%%%%%%%%%%%%%%%%%%%%
As already indicated, if this relation is violated, either (\ref{pi a}) or (\ref{pi b}), or both, are violated.

%%%%%%%%%%%%%%%%%%%%%%
\subsection{Two particles}
\label{schwinger-2-parts.}

Let us now consider the two-particle case, which is the one of interest to us.  Each particle is described in an $N$-dimensional Hilbert space.   We shall use Schwinger unitary operators defined for each particle, and relations similar to Eqs.~(\ref{X(p)}) and (\ref{Z(q)}) of \ref{schwinger-one-particle} to introduce the operators $\hat{p}_i$ and $\hat{q}_i$ which play the role of ``momentum" and ``position" for particle $i$.  \ref{property of P(p1,p2)} shows that under partial transposition of particle 1, for $N>2$ 
(we recall that for $N=2$, $| p\rangle = |-p\rangle$), the joint probability distribution (jpd) of the two momenta is affected as follows:
%%%%%%%%%%%%%%%%%
\begin{equation}  \label{P(p1,p2)}
{\cal P}_{\hat{\rho}^{T_1}}(p_1, p_2)
= {\cal P}_{\hat{\rho}}(-p_1, p_2).
\end{equation}
%%%%%%%%%%%%%%%%%
Thus, PT of particle 1 in the coordinate basis has the intuitive physical meaning  of changing the sign of momentum $p_1$ for particle 1 in the jpd of the two momenta.  The Wigner function, defined as in Refs.~\cite{mello_revzen_2014, mann_mello_revzen_2015}, has the property, shown in \ref{proof of PT and W discrete},
%%%%%%%%%%%%%%%%%%%
\begin{equation}
W_{\hat{\rho}^{T_1}}(q_1,q_2;p_1,p_2)
=W_{\hat{\rho}}(q_1,q_2;-p_1,p_2) \; ,
\label{PT and W discrete}
\end{equation}
%%%%%%%%%%%%%%%%%%%
thus exhibiting again a {\em change in sign of} $p_1$.  
Recall that the definition of the Wigner function of Refs.~\cite{mello_revzen_2014, mann_mello_revzen_2015} requires $N$ to be a prime number larger than 2; see discussion in the previous subsection for one particle.

We define the two-particle operators 
%%%%%%%%%%%%%%%%%%%%%%
\numparts
\begin{eqnarray}
\hat{z} &=& Z_1 Z_2^{\dagger} = \omega^{\hat{q}_1 - \hat{q}_2} \; ,
%\equiv \omega^{\hat{q}},  
%\hspace{1cm} \omega = e^{\frac{2\pi i}{N}}
\label{z'}  \\
\hat{x} &=& X_1 X_2^{\dagger} = \omega^{-(\hat{p}_1 - \hat{p}_2)} \; .
%\equiv \omega^{-2\hat{p}} 
\label{x} 
\end{eqnarray}
\label{z, x}
\endnumparts
%%%%%%%%%%%%%%%%%%%%%%
We emphasize that the operator $\hat{z}$ of Eq.~(\ref{z'}) differs from the ``collective-coordinate" operator defined in Ref.~\cite{revzen2010}, Eq.~(50), which could be written as $\hat{\zeta} \equiv Z_1^{1/2} Z_2^{-1/2}$ and is a variable ``conjugate" to $\hat{x}$ in the sense that it fulfills a commutation relation like that of Eq.~(\ref{comm Z,X}) of \ref{schwinger-one-particle} for one particle.
%We use a boldface character in the definition of the operator 
%$\hat{z}$ of Eq.~(\ref{z'}), to emphasize that it differs from the 
%``collective-coordinate" operator defined in  
%Ref.~\cite{revzen2010}, Eq.~(50), which could be written as $\hat{z}=Z_1^{1/2} Z_2^{-1/2}$,
%which is the variable ``conjugate" to $\hat{x}$ 
%(in the sense that it fulfills a commutation relation like that of Eq.~(\ref{comm Z,X}) for one particle).
As definition (\ref{z'}) does not involve square roots, it does not oblige us, in what follows, to consider prime-number dimensionalities only. 
In terms of these operators we define the Hermitian operators
%%%%%%%%%%%%%%%%%%%%%%
\numparts
\begin{eqnarray}
\hat{A} &=& \frac{1}{2i}(\hat{z}-\hat{z}^{\dagger}) 
= \sin \frac{2 \pi (\hat{q}_1 - \hat{q}_2)}{N}
=\hat{A}^{\dagger} \; ,
\label{A} \\
\hat{B} &=& -\frac{1}{2i}(\hat{x}-\hat{x}^{\dagger})
= \sin \frac{2 \pi (\hat{p}_1 - \hat{p}_2)}{N}
=\hat{B}^{\dagger} \; .
\label{B}
\end{eqnarray}
\label{A,B}
\endnumparts
%%%%%%%%%%%%%%%%%%%%%%
For these Hermitian operators we have the uncertainty relation
%%%%%%%%%%%%%%%%%%%%%%
\begin{equation}
(\Delta \hat{A})_{\hat{\rho}}^2 \; (\Delta \hat{B})_{\hat{\rho}}^2 
\geq \frac14 \left|\left\langle [\hat{A}, \hat{B}] \right\rangle_{\hat{\rho}} \right|^2 \;.
\label{unc-rel 2p 1}
\end{equation}
%%%%%%%%%%%%%%%%%%%%%%
We find that the commutator $[\hat{A}, \hat{B}]$ can be written as
%%%%%%%%%%%%%%%%%%%%%%
\begin{eqnarray}  \label{[A,B] 2p}
[\hat{A}, \hat{B}]
&=& \frac14 
\left\{
[\hat{z}, \hat{x}] - 
[\hat{z}, \hat{x}^{\dagger}] - [\hat{z}^{\dagger}, \hat{x}] + [\hat{z}^{\dagger}, \hat{x}^{\dagger}]
\right\} \nonumber  \\
&=& \frac14
\left[(\omega^2 -1)\left(\hat{x}\hat{z} 
+ \hat{z}^{\dagger} \hat{x}\right)
-((\omega^*)^2 -1)\left(\hat{z}^{\dagger}\hat{x}^{\dagger} + \hat{x}^{\dagger}\hat{z}\right)
\right] \; .
\end{eqnarray}
%%%%%%%%%%%%%%%%%%%%%%
Hence, for a general operator $\hat{\pi}$, we test the uncertainty relation 
%%%%%%%%%%%%%%%%%%%%%%
\begin{eqnarray}
&& (\Delta \hat{A})_{\hat{\pi}}^2 \; (\Delta \hat{B})_{\hat{\pi}}^2 
\geq \frac{1}{64}
\Big|(\omega^2 -1)\left(
\langle \hat{x}\hat{z}\rangle_{\hat{\pi}}
+ \langle \hat{z}^{\dagger}\hat{x}\rangle_{\hat{\pi}} \right)
%\right. 
\nonumber \\
 &&\hspace{3cm} 
 %\left.
-((\omega^{*})^2 -1)
\left(\langle \hat{x}^{\dagger}\hat{z}\rangle_{\hat{\pi}} 
+ \langle\hat{z}^{\dagger}\hat{x}^{\dagger}\rangle_{\hat{\pi}} \right)
\Big|^2 \; .
\label{unc-rel 2p 2}
\end{eqnarray}
%%%%%%%%%%%%%%%%%%%%%%

In \ref{E(A,A2,B,B2,[A,B])} we obtain explicit expressions for the quantities entering this inequality for the case of a general ``density operator" $\hat{\pi}$ which is {\em not necessarily positive definite or Hermitian}.

If relation (\ref{unc-rel 2p 2}) is violated, either (\ref{pi a}) or (\ref{pi b}), or both, are violated.  If $\hat{\pi}$ represents $\hat{\rho}^{T_1}$ and (\ref{unc-rel 2p 2}) is violated, $\hat{\rho}$ is entangled.

%%%%%%%%%%%%%%%%%%%%%%
\section{Applications of the two-particle uncertainty relation}
\label{applications}

%%%%%%%%%%%%%%%%%%%%%%
\subsection{Werner state}
\label{werner}

\begin{figure}[!ht]
\centering
\includegraphics[width=0.5\textwidth]{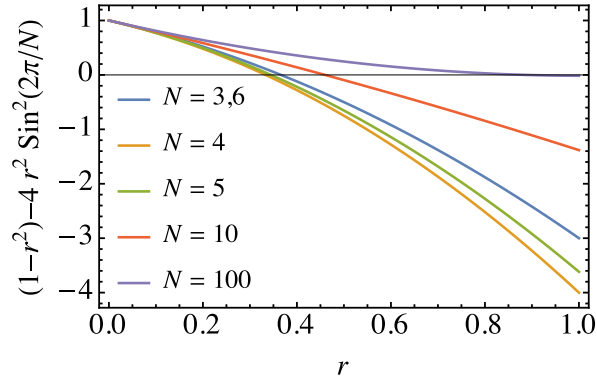}
\caption{(Color online) Plot of the uncertainty relation function $f(r,N) = (1-r)^2 - 4 r^2 \sin^2 \frac{2 \pi}{N}$ versus $r$ for various $N$.  Note that $f(r,N)$ is non-monotonic with $N$: $f(r,N)$ with $N = 3$ is equal to that with $N = 6$, $f(r,N)$ for $N = 4$ is lower than for $N = 5$, which is in turn lower than that for $N = 3, 6$.  The uncertainty relation (\ref{unc-rel 2p 2}) is violated when $f(r,N) < 0$.  The lowest curve is for $N = 4$, the next lowest for $N = 5$, the next for $N = 3$ and 6, the next for $N = 10$ and the highest for $N = 100$.}
\label{Fig_1}
\end{figure}

Consider the Werner state \cite{Werner1989},
%%%%%%%%%%%%%%%%%%%%%%
\begin{equation}
\hat{\rho}_r = r |\Phi^{+} \rangle \langle \Phi^{+}|
+ (1-r)\frac{\hat{\mathbb{I}}}{N^2} \; ,
\label{werner 1}
\end{equation}
%%%%%%%%%%%%%%%%%%%%%%
where $|\Phi^{+} \rangle$ is the pure state
%%%%%%%%%%%%%%%%%%%%%%
\begin{eqnarray}
 |\Phi^{+} \rangle
&=& \frac{1}{\sqrt{N}} \sum_{q'_1=0}^{N-1}
|q'_1, q'_1 \rangle \; .
\label{Phi+ ket} 
\end{eqnarray}
%%%%%%%%%%%%%%%%%%%%%%
When $N=2$, this is the Bell state 
$|\Phi^{+} \rangle = \frac{1}{\sqrt{2}} ( |00 \rangle + |11 \rangle )$ (see, e.g., Ref.~\cite{band-avishai}, Eq.~(5.25a)), 
or a Braunstein-Mann-Revzen (BMR) state \cite{braunstein-mann-revzen}.  The state (\ref{Phi+ ket}) corresponds to the wave function
%%%%%%%%%%%%%%%%%%%%%%
\begin{eqnarray}  \label{Phi+}
\Phi^{+}(q_1, q_2)
&\equiv & \langle q_1 q_2 |\Phi^{+} \rangle
= \frac{1}{\sqrt{N}}\; \delta_{q_1, q_2} \; .
\label{Phi+ wf} 
\end{eqnarray}
%%%%%%%%%%%%%%%%%%%%%%
This is a {\em maximally-entangled pure state}: the partial density operator of particle 1, obtained by tracing $|\Phi^{+} \rangle \langle \Phi^{+}|$ over particle 2, gives $\hat{I}(1)/N$, meaning that the $N$ states of particle 1 are equally probable, thus conforming with the definition given, e.g., in Ref.~\cite{revzen2010}.  The density matrix elements of the Werner state of Eq.~(\ref{werner 1}) are
%%%%%%%%%%%%%%%%%%%%%%
\begin{equation}
\langle q_1 q_2 |\hat{\rho}_r| q'_1 q'_2  \rangle 
= \frac{r}{N} \delta_{q_1, q_2} \delta_{q'_1, q'_2}
+\frac{1-r}{N^2} \delta_{q_1, q'_1} \delta_{q_2, q'_2} \; ,
\label{sigma-r}
\end{equation}
%%%%%%%%%%%%%%%%%%%%%%
and the matrix elements of the PT $\hat{\rho}^{T_1}_r$ are
%%%%%%%%%%%%%%%%%%%%%%
\begin{equation}
\langle q_1 q_2 |\hat{\rho}^{T_1}_r| q'_1 q'_2  \rangle 
= \frac{r}{N} \delta_{q_1, q'_2} \delta_{q_2, q'_1}
+\frac{1-r}{N^2} \delta_{q_1, q'_1} \delta_{q_2, q'_2} \; .
\label{wernerPT 1}
\end{equation}
%%%%%%%%%%%%%%%%%%%%%%
Note that this $\hat{\rho}^{PT}_r$ is Hermitian; however, it is not be positive definite.  From \ref{E(A,A2,B,B2,[A,B])} we find
%%%%%%%%%%%%%%%%%%%%%%
\numparts
\begin{eqnarray}
\langle \hat{A} \rangle_{\hat{\rho}^{T_1}_r} 
&=& \langle \hat{B} \rangle_{\hat{\rho}^{T_1}_r} 
= 0 \; ,
\label{<A>,<B> sigma-r}  \\
\langle \hat{A}^2 \rangle_{\hat{\rho}^{T_1}_r} 
&=& \langle \hat{B}^2 \rangle_{\hat{\rho}^{T_1}_r}
= \frac{1-r}{2}(1-\delta_{N2})  \; ,
\label{<A2>,<B2> sigma-r}  \\
(\Delta \hat{A})_{\hat{\rho}^{T_1}_r}^2 
&=& (\Delta \hat{B})_{\hat{\rho}^{T_1}_r}^2 
= \frac{1-r}{2} (1-\delta_{N2})  \; ,
\label{var A, var B}   \\
\left\langle \hat{x} \hat{z} \right\rangle_{\hat{\rho}^{T_1}_r} 
&=& \left\langle \hat{z}^{\dagger} \hat{x} \right\rangle_{\hat{\rho}^{T_1}_r} 
= r \, \omega^* \; ,
\label{<xz>,<z+z> sigma-r} \\
\left\langle \hat{x}^{\dagger} \hat{z} \right\rangle_{\hat{\rho}^{T_1}_r} 
&=& \left\langle \hat{z}^{\dagger} \hat{x}^{\dagger} \right\rangle_{\hat{\rho}^{T_1}_r} 
= r \, \omega \; ,
\label{<x+z>,<z+x+>} \\
\left\langle [\hat{A}, \hat{B}] \right\rangle_{\hat{\rho}^{T_1}_r} 
&=& 2 i \, r \sin \frac{2 \pi}{N} \; .
\label{<[A,B]>}
\end{eqnarray}
\label{A,A2,{A,B} sigma-r}
\endnumparts
%%%%%%%%%%%%%%%%%%%%%%
The uncertainty relation (\ref{unc-rel 2p 2}), applied to $\hat{\rho}^{T_1}_r$, requires
%%%%%%%%%%%%%%%%%%%%%%
\begin{equation}
(1-r)^2 (1-\delta_{N2}) - 4 r^2 \sin^2 \frac{2 \pi}{N} \geq 0 \; .
\label{unc-rel 3}
\end{equation}
%%%%%%%%%%%%%%%%%%%%%%
For $N=2$, (\ref{unc-rel 3}) gives $0 \geq 0$, an inconclusive result.  We therefore concentrate on $N \ge 3$.  When 
%%%%%%%%%%%%
\begin{equation}
(1-r)^2 - 4 r^2 \sin^2 \frac{2 \pi}{N} < 0, 
\label{unc-rel viol}
\end{equation}
%%%%%%%%%%%%%%%%%%%%%%
the uncertainty relation (\ref{unc-rel 2p 2}) is violated and the original state $\rho_r$ of Eq.~(\ref{werner 1}) {\em is entangled}.
We find that, when $r_{0} < r \leq 1$, where $r_{0} = \frac{1}{2s+1}$ and $s = \sin \frac{2 \pi}{N}$, {\em the state is entangled}.  For example, for 
%%%%%%%%%%%%%%%%%%%%%%
\numparts
\begin{eqnarray}
N&=&3, \hspace{5mm} s=\sqrt{3}/2,\hspace{10mm}  
r_{0} =1/(\sqrt{3} +1)\approx 0.366   \\
N&=&4, \hspace{5mm} s=1,\hspace{10mm}  
\hspace{8mm}r_{0} =1/3   \\
N&=&6, \hspace{5mm} s=\sqrt{3}/2,\hspace{10mm}  
r_{0} = 1/(\sqrt{3} +1)\approx 0.366 \\
N&=&12, \hspace{5mm} s=1/2,\hspace{10mm}  
\hspace{3mm}r_{0} =1/2   \\
N&=&20, \hspace{4mm} s= 0.31, \hspace{13mm} r_{0} \approx 0.62 \\
N &\gg& 1, \hspace{5mm} 
s \approx 2\pi /N \ll 1, \hspace{3mm}  r_{0} \approx 1- 4\pi /N \; .
\end{eqnarray}
\label{unc-rel 6}
\endnumparts
Figure~\ref{Fig_1} graphically illustrates the uncertainty relation condition.  Whenever $f(r,N) < 0$, the uncertainty relation (\ref{unc-rel 2p 2}) is violated and the Werner state of Eq.~(\ref{werner 1}) is entangled.

%%%%%%%%%%%%%%%%%%%%%%
%%%%%%%%%%%%%%%%%%%%%%

The criterion (\ref{unc-rel viol}) arising from the present UR detects entanglement for $r_0<r<1$, with $r_0$ becoming closer to 1 --- a value which corresponds to the pure maximally entangled state (\ref{Phi+ ket}) --- as the dimensionality $N$ increases.  We remark that this does not mean that for other values of $r$ the state is not entangled.  Our measure establishes that $r$ being in the interval $(r_0, 1)$ is sufficient for the state to be entangled, but not necessary: there might be other values of $r$ for which the state is entangled.

%%%%%%%%%%%%%%%%%%%%%%
\begin{figure}[!ht]
\centering
\includegraphics[width=0.5\textwidth]{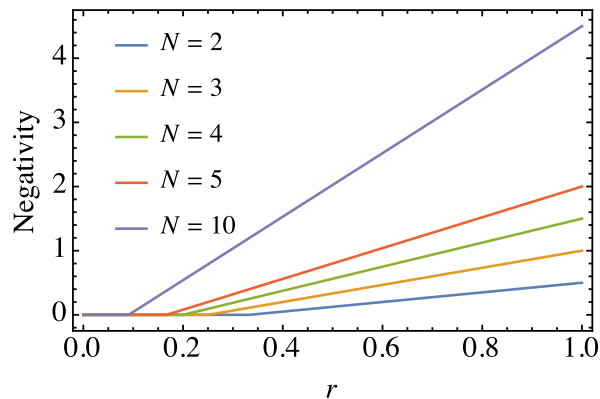}
\caption{(Color online) Plot of the negativity ${\cal N}(r,N) = \frac{1}{2} [\frac{n - 1}{2n} \sqrt{[1 - (n + 1) r]^2} + \frac{n + 1}{2n} \sqrt{[1 + (n - 1) r]^2} - 1]$ versus $r$ for various $N$.  The lowest curve is for $N = 2$ and the curves ${\cal N}(r,N)$ increase with increasing $N$.}
\label{Fig_2}
\end{figure}
%%%%%%%%%%%%%%%%%%%%%%

There might be other measures that are stronger than the present one, in the sense that they may detect entanglement also for other values of $r$ for which the present UR criterion does not. An example we analyzed is provided by the notion of ``negativity" introduced in Ref.~\cite{vidal_werner}.  For $N=3$, our uncertainty relation reveals that the original state is entangled for $0.366 < r < 1$; it does not yield any information for $r < 0.366$.  On the other hand, the notion of negativity predicts entanglement for $0.25 < r < 1$.  Our result is not inconsistent with the negativity prediction. The reason for this difference is that (\ref{unc-rel 3}) is just {\em one} property of a ``proper state'': The uncertainty relation is fulfilled in the range $0.25<r<0.366$, but other properties of a proper state must be violated, in order to break the non-negativity of the ``density matrix" for the partially transposed state.  Figure~\ref{Fig_2} plots the negativity function
%%%%%%%%%%%%
\begin{equation}  \label{Eq:Negativity}
{\cal N}(r,N) = \frac{1}{2} 
\left\{\frac{n - 1}{2n} \sqrt{[1 - (n + 1) r]^2} + \frac{n + 1}{2n} \sqrt{[1 + (n - 1) r]^2} - 1\right\},
\end{equation}
%%%%%%%%%%%%%%
versus $r$ for various values of $N$.  Although Eq.~(\ref{Eq:Negativity}) has been developed here, it conforms to the negativity results of Ref.~\cite{vidal_werner}. The negativity ${\cal N}(r,N)$ is a stronger measure of entanglement than the one arising from the UR. The strength of the negativity measure is even more pronounced for larger values of $N$; the reason is that, as $N$ gets larger, ever more conditions are needed, in addition to the UR, to determine if the state is entangled. We have not examined uncertainty relations involving other observables to see if the entanglement regions shown in Fig.~\ref{Fig_1} are enlarged.

From a conceptual point of view, an attractive feature of the formulation presented in the present paper is that the effect of ${\rm PT_1}$ is intuitively simple to describe: ${\rm PT_1}$ changes the sign of the momentum of particle 1 in the jpd of the two momenta and in the Wigner function (in the latter case for $N$ prime); as a result, an uncertainty relation is violated for certain values of the parameter $r$, from which we conclude that the original state is entangled.  
With the present procedure we are testing entanglement by a measure having a clear physical interpretation,
just as was suggested in the conclusions of Ref.~\cite{vidal_werner}.  

%%%%%%%%%%%%%%%%%%%%%%%%%%
\section{Conclusions}  \label{conclu}

We studied a number of features of bipartate entanglement of discrete dynamical-variable-systems having a  Hilbert space with finite dimensionality $N$ for each particle.  We concentrated on the notion of partial transposition of one of the particles \cite{peres_96,horodecki_97,horodecki_98}; it is well known that a necessary condition for separability is that the density matrix obtained by partial transposition of the density matrix has only nonnegative eigenvalues, i.e., if the ``state" obtained by partial transposition is not a bona fide QM state, the original state is entangled.  We showed that the Wigner function, as formulated in Refs.~\cite{mello_revzen_2014, mann_mello_revzen_2015} (for $N$ prime $>2$), has the intuitive property that under partial transposition the momentum (defined as in these references) of the corresponding particle changes sign [see Eq.~(\ref{PT and W discrete})].  This extends to discrete variables the result found for continuous variables in Ref.~\cite{simon_2000} (see also Refs.~\cite{werner-wolf2001, braunstein_van_loock_2005}).

We formulated an uncertainty relation, Eq.~(\ref{unc-rel 2p 2}), for observables defined in terms of Schwinger operators, for an arbitrary QM state and for an arbitrary dimensionality $N$.  This extends the continuous-variables results of Refs.~\cite{simon_2000, duan_et_al_2000}.  When the uncertainty relation for the partially transposed state is violated, the original state is entangled.  We applied the result to analyze the property of entanglement for a Werner state, constructed as a convex combination with a parameter $r$ of a maximally entangled Braunstein-Mann-Revzen pure state $|\Phi^{+}\rangle\langle \Phi^{+}|$ and the completely incoherent state $\hat{I}/N^2$.  We found the inequality (\ref{unc-rel viol}) which, when fulfilled, signals that the original Werner state is entangled: this occurs for a range of values $r_0 < r < 1$ of the $r$ parameter, where $r_0$ is $N$ dependent.

Reference~\cite{vidal_werner} indicates that quantum correlations and entanglement would be best tested by measures having a physical meaning.  We believe that the interpretation of partial transposition, in conjunction with the uncertainty relation we presented, meet this requirement.

\ack
YBB also acknowledges support by grants from the Israel Science Foundation (Grant No.~295/2011) and the DFG through the DIP program (FO 703/2-1).  PAM acknowledges support by DGAPA, Mexico, under grant IN109014.

%To obtain a simple heading of 'Appendix' use the code

%\section*{Appendix}. 

%If it contains
%numbered equations, figures or tables the command 
%\appendix should precede it and

%\setcounter{section}{1}

%must follow it

\appendix

%%%%%%%%%%%%%%%%%%%%%%%%%%%%%%%%%%
\section{Schwinger operators for one particle}
\label{schwinger-one-particle}

We consider the $N$-dimensional Hilbert space spanned by $N$ distinct states $|q\rangle$, with $q=0,1, \cdots ,(N-1)$, which are subject to the periodic condition $|q+N\rangle=|q\rangle$.  These states are designated as the ``reference basis" of the space.  We follow Schwinger \cite{schwinger} and introduce the unitary operators $\hat{X}$ and $\hat{Z}$, defined by their action on the states of the reference basis by the equations
%%%%%%%%%%%%%%%%%%%%%%%%%%%%%%%
\numparts
\begin{eqnarray}
\hat{Z}|q\rangle
&=&\omega^q \, |q\rangle, \;\;\;\; \omega=e^{2 \pi i/N},
\label{Z}  \\
\hat{X}|q\rangle &=& |q+1\rangle .
\label{X}
\end{eqnarray}
\label{Z,X}
\endnumparts
%%%%%%%%%%%%%%%%%%%%%%%%%%%%%%
The operators $\hat{X}$ and $\hat{Z}$ fulfill the periodicity condition
%%%%%%%%%%%%%%%%%%%%%%%
\begin{equation}
\hat{X}^N = \hat{Z}^N = 
\hat{\mathbb{I}},
\label{X,Z periodic}
\end{equation}
%%%%%%%%%%%%%%%%%%%%%%%%%%%%%%
$\hat{\mathbb{I}}$ being the unit operator.
These definitions lead to the commutation relation
%%%%%%%%%%%%%%%%%%%%%%%%%%%%%%
\begin{equation}
\hat{Z}\hat{X}=\omega \, \hat{X}\hat{Z} .
\label{comm Z,X}
\end{equation}
%%%%%%%%%%%%%%%%%%%%%%%%%%%%%%
The two operators $\hat{Z}$ and $\hat{X}$ form a complete algebraic set, in that only a multiple of the identity commutes with both \cite{schwinger}.  As a consequence, any operator defined in our $N$-dimensional Hilbert space can be written as a function of $\hat{Z}$ and $\hat{X}$.  We also introduce (i.e., define) the Hermitian operators  $\hat{p}$ and $\hat{q}$, which play the role of ``momentum" and ``position", through the equations \cite{de_la_torre-goyeneche, durt_et_al}
%%%%%%%%%%%%%%%%%%%%%%%%%%%%%%%
\numparts
\begin{eqnarray}
\hat{X}
&=& \omega^{-\hat{p}}
= e^{-\frac{2\pi i}{N}\hat{p}} \; ,
\label{X(p)}     \\
\hat{Z}
&=& \omega^{\hat{q}}
=e^{\frac{2\pi i}{N}\hat{q}} \; .
\label{Z(q)} 
\end{eqnarray}
\label{X(p),Z(q)}
\endnumparts
%%%%%%%%%%%%%%%%%%%%%%%%%%%%%%
What we defined as the reference basis can thus be considered as the ``position basis".  With (\ref{comm Z,X}) and definitions (\ref{X(p)}), (\ref{Z(q)}), the commutator of $\hat{q}$ and $\hat{p}$ in the continuous limit 
\cite{de_la_torre-goyeneche,durt_et_al} is the standard one, $[\hat{q},\hat{p}]=i$. 

%%%%%%%%%%%%%%%%%%%%%%%%%%%%%%
\section{Proof of Eq. (\ref{P(p1,p2)})}
\label{property of P(p1,p2)}

The jpd of the two momenta $p_1, p_2$ in the state $\hat{\rho}$ is given by
%%%%%%%%%%%%%%%%%%%%%%
\numparts
\begin{eqnarray}
{\cal P}_{\hat{\rho}}(p_1,p_2)
&=& {\mathrm{Tr}} (\hat{\rho} \; \mathbb{P}_{p_1} \otimes \mathbb{P}_{p_2}) 
\label{P(p1,p2) a}   \\
&=&\sum_{n_1 n_2 n_1'n_2'}
\langle n_1, n_2 | \hat{\rho} | n_1', n_2' \rangle
\langle n_1', n_2'| \mathbb{P}_{p_1} \otimes \mathbb{P}_{p_2}|n_1, n_2 \rangle   
\label{P(p1,p2) b}   \\
&=&\frac{1}{N^2}\sum_{n_1 n_2 n_1'n_2'} 
\langle n_1, n_2 | \hat{\rho} | n_1', n_2' \rangle
\omega^{p_1(n_1'-n_1)} 
\omega^{p_2(n_2'-n_2)} .
\label{P(p1,p2) c}      
\end{eqnarray}
\label{}
\endnumparts
%%%%%%%%%%%%%%%%%%%%%%
The jpd of the two momenta $p_1, p_2$ for the PT operator $\hat{\rho}^{T_1}$ is given by
%%%%%%%%%%%%%%%%%%%%%%
\numparts
\begin{eqnarray}
{\cal P}_{\hat{\rho}^{T_1}}(p_1,p_2)
&=& {\mathrm{Tr}} (\hat{\rho}^{T_1} \mathbb{P}_{p_1} \otimes \mathbb{P}_{p_2}) 
\label{P(p1,p2) a}   \\
&=&\sum_{n_1 n_2 n_1'n_2'}
\langle n_1, n_2 | \hat{\rho}^{T_1} | n_1', n_2' \rangle
\langle n_1', n_2'| \mathbb{P}_{p_1} \otimes \mathbb{P}_{p_2}|n_1, n_2 \rangle  
\label{P(p1,p2) b}   \\
&=&\frac{1}{N^2}\sum_{n_1 n_2 n_1'n_2'} 
\langle n_1, n_2 | \hat{\rho} | n_1', n_2' \rangle
\omega^{-p_1(n_1'-n_1)} 
\omega^{p_2(n_2'-n_2)} 
\label{P(p1,p2) c}   \\
&=& {\cal P}_{\hat{\rho}}(-p_1,p_2) .
\end{eqnarray}
\label{}
\endnumparts
%%%%%%%%%%%%%%%%%%%%%%
This proves Eq. (\ref{P(p1,p2)}).
The above proof applies for $N>2$, since, for $N=2$, $| p\rangle = |-p\rangle$.

For the case of only one particle, the above result reduces to that of Eq. (\ref{P(p) 1part}).

%%%%%%%%%%%%%%%%%%%%%%%%%%%%%%
\section{Proof of Eq. (\ref{PT and W discrete})}
\label{proof of PT and W discrete}

We define the Wigner function for the density operator $\hat{\rho}$ as in Refs.~\cite{mello_revzen_2014, mann_mello_revzen_2015, revzen_epl_2012}, as
%%%%%%%%%%%%%%%%%%%
\begin{equation}  \label{WF discrete 1}
W_{\hat{\rho}}(q_1,q_2,p_1,p_2) = {\mathrm{Tr}}\left[\hat{\rho}(\hat{P}_{q_1 p_1}\otimes \hat{P}_{q_2 p_2})\right] \; ,
\end{equation}
%%%%%%%%%%%%%%%%%%%
where $\hat{P}_{q_i p_i}$ is the ``line operator" for particle $i$, defined in Refs.~\cite{mello_revzen_2014, mann_mello_revzen_2015, revzen_epl_2012}.  Explicitly, we find
%%%%%%%%%%%%%%%%%%%
\begin{equation}  \label{WF discrete 2}
W_{\hat{\rho}}(q_1,q_2,p_1,p_2) = \sum_{q_1' q_2'} \langle q_1', q_2'| \hat{\rho}  | 2q_1-q_1', 2q_2-q_2' \rangle \omega^{2p_1 (q_1-q_1')} \omega^{2p_2 (q_2-q_2')} .
\end{equation}
%%%%%%%%%%%%%%%%%%%

By definition, the Wigner function after ${\rm PT_1}$ is then
%%%%%%%%%%%%%%%%%%%%%%
\numparts
\begin{eqnarray}
W_{\hat{\rho}^{T_1}}(q_1,q_2,p_1,p_2)
&=& \sum_{q_1' q_2'}
\langle 2q_1-q_1', q_2'| \hat{\rho}  |q_1', 2q_2-q_2'  
\rangle \omega^{2p_1 (q_1-q_1')} \omega^{2p_2 (q_2-q_2')} 
\label{WF PT discrete a}
\nonumber \\ \\
&=& \sum_{q_1' q_2'}
\langle q_1'', q_2'| \hat{\rho}  |2q_1-q_1'', 2q_2-q_2'  
\rangle \omega^{2(-p_1) (q_1-q_1'')} \omega^{2p_2 (q_2-q_2')} 
\label{WF PT discrete b}   \nonumber \\ \\
&=&W_{\hat{\rho}}(q_1,q_2,-p_1,p_2) \; ,
\end{eqnarray}
\label{}
\endnumparts
%%%%%%%%%%%%%%%%%%%%%%
where we made the change of variables $2q_1-q_1'=q_1''$.  This proves Eq.~(\ref{PT and W discrete}).

For the case of only one particle, the above result reduces to that of Eq. (\ref{PT and W discrete 1part}).

%%%%%%%%%%%%%%%%%%%%%%%%%%%%%%
\section{The quantities entering the uncertainty relation (\ref{unc-rel 2p 2})}
\label{E(A,A2,B,B2,[A,B])}

We write explicitly the quantities entering the uncertainty relation 
(\ref{unc-rel 2p 2}) for the case of a general ``density operator" $\hat{\pi}$, 
{\em not necessarily positive definite or Hermitian}:
%%%%%%%%%%%%%%%%%%%%%%
\numparts
\begin{eqnarray}
\langle \hat{A} \rangle_{\hat{\pi}}
&=& \sum_{q_1,q_2}\langle q_1 q_2|\hat{\pi} |q_1 q_2 \rangle 
\sin \frac{2\pi(q_1 - q_2)}{N} \; ,
\label{<A>} \\
\langle \hat{A^2} \rangle_{\hat{\pi}} 
&=& \sum_{q_1,q_2}\langle q_1 q_2|\hat{\pi} |q_1 q_2 \rangle 
\sin^2 \frac{2\pi(q_1 - q_2)}{N} \; ,
\label{<A2>} \\
\langle \hat{B} \rangle_{\hat{\pi}}
&=& \frac{i}{2}\sum_{q_1,q_2}
\left[\left\langle q_1 q_2|\hat{\pi} |q_1+1, q_2-1 \right\rangle 
-\left\langle q_1 q_2|\hat{\pi} |q_1-1, q_2+1 \right\rangle \right] \; ,
\label{<B>} \\
\langle \hat{B}^2 \rangle_{\hat{\pi}}
&=& \frac{1}{2} {\mathrm{Tr}} \, \hat{\pi}
-\frac{1}{4}\sum_{q_1,q_2}
\left[\left\langle q_1 q_2|\hat{\pi} |q_1+2, q_2-2 \right\rangle 
+ \left\langle q_1 q_2|\hat{\pi} |q_1-2, q_2+2 \right\rangle \right] \; ,
\nonumber \\
\label{<B2>} 
\end{eqnarray}
\label{<A>,...,<B2>}
\endnumparts
%%%%%%%%%%%%%%%%%%%%%%
and
%%%%%%%%%%%%%%%%%%%%%%
\numparts
\begin{eqnarray}
\langle \hat{x} \hat{z} \rangle_{\hat{\pi}}
&=& \sum_{q_1,q_2}
\left\langle q_1 q_2|\hat{\pi} |q_1+1, q_2-1 \right\rangle 
\omega^{q_1-q_2} \; ,
\label{<xz>} \\
\langle \hat{z}^{\dagger} \hat{x} \rangle_{\hat{\pi}}
&=& \sum_{q_1,q_2}
\left\langle q_1 q_2|\hat{\pi} |q_1+1, q_2-1 \right\rangle 
\omega^{q_2-q_1-2} \; ,
\label{<z+x>} \\
\langle \hat{x}^{\dagger} \hat{z} \rangle_{\hat{\pi}}
&=& \sum_{q_1,q_2}
\left\langle q_1 q_2|\hat{\pi} |q_1-1, q_2+1 \right\rangle 
\omega^{q_1-q_2} \; ,
\label{<x+z>}  \\
\langle \hat{z}^{\dagger} \hat{x}^{\dagger} \rangle_{\hat{\pi}}
&=& \sum_{q_1,q_2}
\left\langle q_1 q_2|\hat{\pi} |q_1-1, q_2+1 \right\rangle 
\omega^{-q_1+q_2+2} \; .
\label{<z+x+>}
\end{eqnarray}
\label{<XZ>,<Z+X>}
\endnumparts
%%%%%%%%%%%%%%%%%%%%%%

%%%%%%%%%%%%%%%%%%%%%%%%%%%%%%%%%%


\section*{Bibliography}
\begin{thebibliography}{99}

\bibitem{Steane_98}
A. Steane, 
%``Quantum computing'',
Rep. Prog. Phys. {\bf 61}, 117,  (1998). %Ð173.

\bibitem{band-avishai}
Y. B. Band and Y. Avishai,
{\em Quantum Mechanics}, (Academic Press (Elsevier), Oxford, 2013).

\bibitem{plenio_virmani_2006}
M. B. Plenio and S. Virmani, 
%``An introduction to entanglement measures'',
Quant. Inf. Comput. {\bf 7}, 1-51 (2007).
%arXiv:0504163v3 [quant-ph] (2006).

\bibitem{peres_96}
A. Peres, 
Phys. Rev. Lett. {\bf 77}, 1413 (1996).

\bibitem{horodecki_97}
P. Horodecki,
Phys. Lett. A {\bf 232}, 333 (1997).

\bibitem{horodecki_98}
M. Horodecki P. Horodecki and R. Horodecki,
Phys. Rev. Lett. {\bf 80}, 5239 (1998).

\bibitem{simon_2000}
R. Simon,
Phys. Rev. Lett. {\bf 84}, 2726 (2000).

\bibitem{werner-wolf2001}
R. F. Werner and M. M. Wolf,
Phys. Rev. Lett. {\bf 86}, 3658 (2001).

\bibitem{braunstein_van_loock_2005}
S. L. Braunstein and P. van Loock,
Rev. Mod. Phys. {\bf 77}, 513 (2005).

\bibitem{duan_et_al_2000}
L-M Duan, G. Giedke, J. J. Corac. and P. Zoller,
Phys. Rev. Lett. {\bf 84}, 2722 (2000).

\bibitem{mello_revzen_2014}
P. A. Mello and M. Revzen,
Phys. Rev. A {\bf 89}, 012106 (2014).

\bibitem{mann_mello_revzen_2015}
A. Mann, P. A. Mello and M. Revzen,
arXiv:1507.03167 [quant-ph] (2015).

\bibitem{schwinger} 
J. Schwinger, 
%``Unitary Operator Bases",
Proc. Nat. Acad. Sci. (USA) {\bf 46}, 570-579 (1960);
%``Unitary transformations and the action principle",
Ibid,
%Proc. Nat. Acad. Sci. (USA) {\bf 46}, 
883-897 (1960).

\bibitem{de_la_torre-goyeneche}
A. C. de la Torre and D. Goyeneche,
Am. J. Phys. {\bf 71}, 49 (2003).

\bibitem{durt_et_al}
T. Durt, B-G Englert, I. Bebgtsson and K. Zyczkowski,
Int. Jour. Quant. Inf. {\bf 8}. 535 (2010).

\bibitem{wootters87}
W. K. Wootters,
Ann. Phys. (N.Y.) {\bf 176}, 1 (1987).

\bibitem{wooters-fields89}
W. K. Wootters and B. D. Fields,
Ann. Phys. (N.Y.), {\bf 191}, 363 (1989)

\bibitem{ivanovic36}
I. D. Ivanovic, J. Phys. A {\bf 14}, 3241 (1981).

\bibitem{ballentine}
L. E. Ballentine,
{\em Quantum Mechanics: A Modern Development}, 
(World Scientific Publishing Co., Singapore, 1998).

\bibitem{revzen2010}
M. Revzen,
Phys. Rev. A{\bf 81}, 012113 (2010).

\bibitem{Werner1989}
R. F. Werner, 
%``Quantum states with Einstein-Podolsky-Rosen correlations admitting a hidden-variable model'', 
Phys. Rev. A{\bf 40}, 4277-4281, (1989).

\bibitem{braunstein-mann-revzen}
S. L. Braunstein, A. Mann, and M. Revzen,
Phys. Rev. Lett. {\bf 68}, 3259 (1992).

\bibitem{vidal_werner}
G. Vidal and R. F. Werner,
Phys. Rev. A {\bf 65}, 032314 (2002).

\bibitem{revzen_epl_2012}
M. Revzen, Europh. Lett. {\bf 98}, 10001 (2012).

\end{thebibliography}
\end{document}